\documentclass[aip,jcp,reprint,groupedaddress,floatfix]{revtex4-1}
\pdfoutput=1

\usepackage{siunitx} 

\usepackage{amsmath} 
\usepackage{amssymb}
\usepackage{mathrsfs}
\usepackage{mathtools}
\usepackage{textcomp}
\usepackage{stmaryrd}
\usepackage{esint}

\usepackage[shortlabels]{enumitem}
\usepackage{graphicx} 

\usepackage{multirow}
\usepackage{booktabs} 
\usepackage{hyperref} 

\newcommand\abs[1]{\left|#1\right|}

\begin{document}


\title{Hierarchical bounding structures for efficient virial computations: Towards a realistic molecular description of cholesterics}
\date{\today}
\author{Maxime M.C. Tortora}
\author{Jonathan P.K. Doye}
\affiliation{Physical and Theoretical Chemistry Laboratory, University of Oxford, South Parks Road, Oxford, OX1 3QZ, United Kingdom}



\begin{abstract}
We detail the application of bounding volume hierarchies to accelerate second-virial evaluations for arbitrary complex particles interacting through hard and soft finite-range potentials. This procedure, based on the construction of neighbour lists through the combined use of recursive atom-decomposition techniques and binary overlap search schemes, is shown to scale sub-logarithmically with particle resolution in the case of molecular systems with high aspect ratios. Its implementation within an efficient numerical and theoretical framework based on classical density functional theory enables us to investigate the cholesteric self-assembly of a wide range of experimentally-relevant particle models. We illustrate the method through the determination of the cholesteric behaviour of hard, structurally-resolved twisted cuboids, and report quantitative evidence of the long-predicted phase handedness inversion with increasing particle thread angles near the phenomenological threshold value of $45^\circ$. Our results further highlight the complex relationship between microscopic structure and helical twisting power in such model systems, which may be attributed to subtle geometric variations of their chiral excluded-volume manifold.
\end{abstract}


\maketitle 


\section{Introduction} \label{sec:Introduction}
Liquid crystals (LCs) constitute an intriguing class of structured fluids, whose hallmark characteristic lies in the intricate link between macroscopic organisation and the microscopic properties of their constituent particles.\cite{deGennes} This delicate combination of long-range order and mechanical fluidity allows for the reversible tuning of their phase structure through the fine application of external fields and thermochemical stimuli, leading to a wealth of practical uses. Common examples include the ubiquitous field-switching technologies underlying all modern LC displays, but also applications as templates for the design of micro- and nano-structured materials and as defect-free semiconductors for organic electronics and energy conversion purposes.\cite{Lagerwall-1}
\par
The subtle dependence of their long-range organisation on detailed microscopic structure renders the quantitative prediction of LC assemblies a formidable task, owing to the vast discrepancies between the typical length- and time-scales of local molecular motion and those associated with the macroscopic phase-ordering process.\cite{Wilson-5} The reliable numerical investigation of many experimentally-relevant systems through atomistic simulation methods thus usually requires the combined use of complex, accurate force fields and large simulation boxes over long equilibration times, often leading to considerable computational costs.\cite{Wilson-1, Wilson-3} While significant speed-ups can be obtained through the careful \textit{ab-initio} parametrization of coarse-grained molecular potentials,\cite{Cacelli} such schemes are however still hindered by the high numerical expense ensuing from the evaluation of non-bonded interaction terms, which underpin the self-assembly of LCs.
\par
A notable illustration of these limitations is the case of the \textit{cholesteric phase}, formed by many systems of chiral particles, in which the local direction of alignment of the molecules --- termed the \textit{nematic director} --- periodically rotates around a given axis in space.\cite{deGennes} The fluid-like positional disorder typically associated with this level of orientational organisation allows for the bulk structure of such phases to be fully characterised in terms of a single scalar $\mathcal{P}$, known as the \textit{cholesteric pitch}, corresponding to the spatial period of the director rotation. This fascinating display of structural helicity in a liquid phase endows cholesterics with unique optical and mechanical properties, which have been exploited in an impressive variety of applications over the last decades.\cite{Meier, Woltman}
\par
Despite their relatively simple symmetry and singularly long history,\cite{Reinitzer} the general link between microscopic chirality and macroscopic organisation in cholesterics has remained largely elusive. Considerable efforts have been devoted to investigating the effects of particle properties and chemical conditions on liquid crystalline behaviour for a variety of synthetic and bio-polymer solutions, ranging from filamentous viruses\cite{Dogic-1, Dogic-2, Grelet} and DNA duplexes\cite{vanWinkle, Livolant, Stanley} to cellulose\cite{Werbowyj,Revol-1} and chitin nano-crystals.\cite{Revol-2,Murray} The intricacy of the pitch dependence on molecular structure, concentration, temperature and solvent effects has been extensively reported in a number of such colloidal LCs in recent years.\cite{Sato,Tomar}
\par
In contrast with this complex experimental backdrop, most theoretical studies of cholesterics have so far been mainly focused on highly-idealised model systems,\cite{Osipov, Odijk-2,Evans,Pelcovits,Varga-1,Wensink-1,Wensink-2,Wensink-3,Ferrarini-4,Dussi-1,Dussi-2} while simulation studies have been restricted by computational limitations to simple particle models in the regime of low aspect ratios and short cholesteric pitches.\cite{Masters,Zannoni,Memmer,Varga-2,Dussi-3,Wensink-4,Schilling} Rare attempts to directly address experimental systems, including DNA\cite{Podgornik,Kornyshev-1,Kornyshev-2,Kornyshev-3,Ferrarini-1} and filamentous viruses,\cite{Ferrarini-2} have been limited to rigid, coarse-grained representations of the particles, and had to resort to a number of further numerical approximations to reduce the computational burden.\cite{Ferrarini-1}
\par
We have recently introduced an extensive numerical framework based on classical density-functional theory\cite{Onsager,Straley-2} (DFT) to efficiently obtain the bulk equilibrium properties of nematic and cholesteric phases from the microscopic structure and interaction potential of their constituent mesogens.\cite{Tortora} We here propose to combine this theoretical description with an original high-performance virial integration scheme, employing state-of-the-art recursive acceleration structures to expedite non-bonded pair energy evaluations for systems of high-aspect-ratio particles with arbitrary levels of structural complexity. The conjunction of these two developments provides an efficient workaround to bridge the gap between simulations and experiments in molecular investigations of cholesteric self-assembly, free from further numerical or analytical constraints.
\par
The structure of the paper is organised as follows. We first briefly summarise in Sec.~\ref{sec:Perturbative density functional theory for cholesteric phases} the main features of the theoretical approach detailed in Ref.~\onlinecite{Tortora} for the microscopic description of cholesteric mesophases. We then present in Sec.~\ref{sec:Recursive acceleration structures for virial integration} a neighbour list algorithm based on hierarchical bounding volumes for the efficient computation of the generalised virial coefficients underlying our molecular field theory, and proceed to illustrate in Sec.~\ref{sec:The cholesteric behaviour of hard twisted cuboids} the results and performance of its application to systems of hard twisted cuboidal mesogens with various geometric shapes and discretized surfaces. We finally conclude in Sec.~\ref{sec:Conclusion and outlook} with a few remarks on the general link between particle chirality and cholesteric structure in such systems, and highlight some further prospective uses of our theoretical framework and numerical implementation.

\section{Perturbative density functional theory for cholesteric phases} \label{sec:Perturbative density functional theory for cholesteric phases}
The theoretical description outlined here is equivalent to the approach introduced by Straley,\cite{Straley-2} and constitutes a natural extension of the seminal Onsager DFT\cite{Onsager} to bulk nematic phases in the limit of weak director fluctuations. The local structure of a generic uniaxial nematic system may in this formalism be fully described in terms of the single-particle density field,
\begin{equation}
  \label{eq:density_onsager}
  \rho(\mathbf{r}, \mathbf{u}) = \rho \times \psi\left[\mathbf{n}(\mathbf{r})\cdot \mathbf{u}\right],
\end{equation}
where $\rho \equiv N/V$ is the uniform number density and $\psi$ is the orientation distribution function (ODF) quantifying the probability of finding a particle with long axis $\mathbf{u}$ at position $\mathbf{r}$ with respect to the local nematic director $\mathbf{n}(\mathbf{r})$.
\par
In this framework, a helical cholesteric assembly of wavenumber $q=2\pi/\mathcal{P} \ll 1$ may be considered as a weak distortion with respect to a reference nematic state of uniform director field $\mathbf{n}\equiv\mathbf{e}_z$. The free energy $\mathscr{F}_q$ of such a phase may then be conveniently expanded up to quadratic order in powers of $q$,\cite{Wensink-1,Wensink-3}
\begin{equation}
\begin{split}
  \label{eq:free_energy}
  \frac{\beta\mathscr{F}_q}{V} =& \:4\pi^2 \rho \int_0^\pi d\theta \times \sin \theta \psi(\cos\theta) \Big\{\log [\rho \psi(\cos\theta)]-1 \Big\} \\ &+ \beta \left( \kappa_{00} - \kappa_{01} q + \kappa_{11} \frac{q^2}{2} \right),
\end{split}
\end{equation}
where the first term represents the free energy of an ideal gas of anisotropic particles with angular distribution $\psi(\cos\theta)\equiv\psi(\mathbf{u}\cdot \mathbf{e}_z)$ at inverse temperature $\beta$.
\par
The thermodynamic dependence of $\mathscr{F}_q$ on the spatial modulations of the director field is thus purely described by the $\kappa_{ij}$ coefficients, which quantify the deviations from ideal behavior stemming from local intermolecular forces. Denoting by $U(\mathbf{r}_{12}, \mathcal{R}_1, \mathcal{R}_2)$ the interaction energy between a pair of particles with relative center-of-mass separation $\mathbf{r}_{12}$ and respective orientations $\mathcal{R}_1$, $\mathcal{R}_2$, these quantities may be cast under the compact form\cite{Tortora}
\begin{equation}
  \label{eq:elastic_cst}
  \begin{split}
  \beta\kappa_{ij} =& -\frac{\rho^2}{2} \int_V d\mathbf{r}_{12} \oiint d\mathcal{R}_1 d\mathcal{R}_2 \times f(\mathbf{r}_{12},  \mathcal{R}_1,  \mathcal{R}_2) \\ & \times \psi^{(i)}(u_1^z)\Big(-u_1^y r_{12}^x\Big)^i \times \psi^{(j)}(u_2^z)\Big(u_2^y r_{12}^x\Big)^j,
  \end{split}
\end{equation}
with $\mathbf{e}_x$ the cholesteric helical axis, $v^k \equiv \mathbf{v} \cdot \mathbf{e}_k$ for any vector $\mathbf{v}$ and $\psi^{(i)}$ the $i$-th derivative of $\psi$. In the context of the second-virial approximation, the excluded volume kernel $f$ simply corresponds to the Mayer $f$-function,\cite{Tortora}
\begin{equation}
  \label{eq:mayer}
  f(\mathbf{r}_{12},  \mathcal{R}_1,  \mathcal{R}_2) = \exp\Big[-\beta U(\mathbf{r}_{12}, \mathcal{R}_1, \mathcal{R}_2)\Big] - 1,
\end{equation}
and is exact in the limit of high particle aspect ratios.\cite{Parsons,Lee}
\par
The perturbative introduction of chirality in the framework of the Onsager theory revolves around the assumption that local phase structure is largely unaffected by long-range nematic fluctuations, and that molecular arrangements about the local director are therefore not contingent on the field topology of the latter.\cite{Straley-1,Straley-2} The significance of this approximation has been extensively discussed in Ref.~\onlinecite{Tortora}, in which its quantitative validity was demonstrated for cholesteric pitches as short as a few dozen particle diameters.
\par
It then follows from Eqs.~\eqref{eq:density_onsager}--\eqref{eq:elastic_cst} that the equilibrium properties of a cholesteric phase at fixed density $\rho$ are conditioned by the ODF $\psi_{\rm eq}^{(\rho)}$ of its uniform nematic counterpart, which may be directly determined by functional minimisation of the corresponding free energy $\mathscr{F}_0$. A tractable expression for $\psi_{\rm eq}^{(\rho)}$ may be obtained in terms of the angle-dependent second-virial coefficient
\begin{equation}
  \label{eq:second_virial}
  \begin{split}
  E(\theta,\theta') = &-\int_V d\mathbf{r}_{12} \oiint d\mathcal{R}_1 d\mathcal{R}_2 \times f(\mathbf{r}_{12}, \mathcal{R}_1, \mathcal{R}_2) \\ &\times \delta\Big(\cos \theta -u_1^z\Big) \delta\Big(\cos \theta' -u_2^z\Big),
  \end{split}
\end{equation}
where $\delta$ denotes the Dirac distribution, and can be cast in the form of the self-consistent equation\cite{Onsager}
\begin{equation}
  \begin{split}
  \label{eq:self-consistent}
  \psi^{(\rho)}_{\rm eq}(\cos\theta) = &\:\frac{1}{\lambda} \times \exp\bigg[-\frac{\rho}{4\pi^2} \int_0^\pi d\theta'  \\ &\times \sin\theta' E(\theta, \theta') \psi^{(\rho)}_{\rm eq}(\cos\theta') \bigg],                                           
  \end{split}
\end{equation}
with $\lambda$ a Lagrange multiplier ensuring the normalisation of the angular distribution. 
\par
A simple physical interpretation of the $\kappa_{ij}$ coefficients may then be derived by analogy with the Oseen-Frank functional, which provides a mean-field relation between free energy fluctuations and director gradients in weakly non-uniform nematic phases.\cite{Frank} In this formalism, the quadratic term $\kappa_{11}$ of the cholesteric free energy Eq.~\eqref{eq:free_energy} --- quantifying the thermodynamic penalty associated with helical director distortions of either handedness --- may be identified as the so-called \textit{Frank twist elastic constant}, while the linear term $\kappa_{01}$ corresponds to the \textit{chiral strength} promoting a non-zero phase twist to accommodate local particle chirality.\cite{deGennes} The equilibrium cholesteric wavenumber $q_{\rm eq}$ of the system minimising $\mathscr{F}_q$ at fixed density $\rho$ thus results from the exact compensation of these two antagonistic torques,
\begin{equation}
  \label{eq:pitch}
  q_{\rm eq}(\rho) = \frac{\kappa_{01}[\psi^{(\rho)}_{\rm eq}]}{\kappa_{11}[\psi^{(\rho)}_{\rm eq}]} \equiv \frac{2\pi}{\mathcal{P}_{\rm eq}(\rho)},
\end{equation}
and may be evaluated by solving Eq.~\eqref{eq:self-consistent} iteratively\cite{Herzfeld} for the chosen particle model and plugging the resulting numerical ODF into Eq.~\eqref{eq:elastic_cst}.
\par
Hence, Eqs.~\eqref{eq:density_onsager}--\eqref{eq:pitch} reduce the determination of equilibrium phase structure to the computation of generalised virial integrals of the form Eqs.~\eqref{eq:elastic_cst} and \eqref{eq:second_virial} for systems of arbitrary mesogens in the limit of high particles aspect ratios and long cholesteric pitches. Such integrals may be conveniently performed by Monte-Carlo (MC) sampling,\cite{Metropolis} which provides an efficient and scalable framework to explore the two-particle configuration space of a wide range of molecular systems interacting through hard or soft potentials,\cite{Dussi-1,Dussi-2} and form the microscopic basis of our field-theoretical description.\footnote{Note that, as in Ref.~\onlinecite{Tortora}, we have here made use of the Parsons-Lee approximation,\cite{Parsons,Lee} but have omitted the corresponding prefactor from Eqs.~\eqref{eq:elastic_cst}--\eqref{eq:self-consistent} due to its small effect for the high particle aspect ratios considered in this paper.}

\section{Recursive acceleration structures for virial integration} \label{sec:Recursive acceleration structures for virial integration}
The main performance bottleneck of this approach lies in the computation of the particle pair interaction energy $U$ as part of the stochastic sampling of Eqs.~\eqref{eq:elastic_cst}--\eqref{eq:second_virial}. In common with most molecular modelling methods, the origin of this numerical expense largely resides in the number of inter-atom\footnote{Note that we here use the term \textit{atoms} rather loosely to refer to the elementary building blocks comprising a single particle. We further use the terms ``particle'' and ``molecule'' interchangeably to denote a full mesogenic compound.} distances to be computed at every step for the evaluation of the various non-bonded interaction potentials comprising $U$. In the pairwise additive approximation, the naive time complexity of molecular simulations is indeed generally quadratic in the number of their constituent atoms, leading to rapidly prohibitive computational costs with increasing system sizes and particle complexity. 
\par
While the use of density functional methods enables us to considerably alleviate this numerical burden by providing a statistical-mechanical link between bulk equilibrium properties and pair molecular interactions, the high statistical accuracy required in the numerical evaluation of Eqs.~\eqref{eq:elastic_cst} and \eqref{eq:pitch} for the reliable prediction of long cholesteric pitches entails the use of a large number of MC steps to ensure the full convergence of these virial integrals. The corresponding number of pair interaction energies to be computed may thus still hinder the tractable treatment of many experimentally-realistic systems, and one must therefore devise general numerical schemes for the efficient calculation of inter-particle potentials in the case of complex atomistic particle models.
\par
A traditional workaround, as employed in most modern simulation frameworks, is to exploit the locality of finite-range interaction potentials through the use of cell lists,\cite{Allen} which generally enable one to achieve a linear time complexity in the number of particles by restricting energy evaluation queries to pairs of neighbouring atoms. These methods usually rely on the spatial partition of the simulation box into a grid of cells, into which the different constituent particles may be binned in order to expedite the determination of all atom pairs within the interaction cutoff range. However, the memory scaling of these simple acceleration structures with system volume combined with their ineffective handling of anisotropic interaction potentials\cite{Rovigatti} render them ill-suited for the study of disperse solutions of high aspect ratio particles --- as in the case of many colloidal liquid crystals.
\par
Recent numerical studies\cite{Glotzer-1, Glotzer-2} have proposed to circumvent these issues through the construction of neighbour lists based on the partitioning of the particles themselves, rather than that of their total accessible volume. These algorithms, inspired by general considerations borrowed from the field of graphics rendering, instead resort to the use of recursive data structures known as bounding volume hierarchies (BVHs) to accelerate the binning of neighbouring atoms into simple geometrical primitives. Their parallel implementation on graphical processing units (GPUs) was thus found to yield sizeable performance gains over standard grid-based methods in molecular dynamics simulations of non-uniform colloidal solutions with high size polydispersity.\cite{Glotzer-1}
\par
BVHs are commonly used in many computational geometry applications in which the rapid detection of potential collisions involving complex moving objects is critical; examples include motion planning in robotics,\cite{Rock} ray casting in real-time graphics rendering\cite{Shirley} and the physics engine of both video games\cite{Eberly} and computer-aided design platforms.\cite{Ding} The popularity of BVHs stems from their ability to capture the geometric features of a wide variety of 3D objects at successive levels of spatial resolution, and hence quickly prune the sections of their structure contained in non-overlapping bounding volumes, as illustrated in Figs.~\ref{fig1} and \ref{fig2}. Similar culling procedures have been successfully employed in recent years for the efficient implementation of neighbour search in systems of complex molecules, both rigid\cite{Grudinin-1, Grudinin-2} and flexible.\cite{Latombe}
\par
One of the challenges in designing effective BVHs lies in the choice of the elementary bounding volumes to be used, as their performance is generally contingent on two antagonistic factors. Namely, (i) the bounding volumes should fit the chosen particle model as tightly as possible in order to minimise the number of distance queries to be performed at the end of the pruning process, and (ii) the computational cost of testing two bounding volumes for overlap should be minimal, so as to ensure the rapid traversal of the BVH.\cite{Ericson} Therefore, there exists no universal choice of a BVH yielding optimal performance in all situations, and one must rather tailor the selection of a geometrical primitive to the particular system studied, keeping the two previous considerations in mind.\cite{Klosowski}
\par
A good overall compromise between tightness and computational expense was achieved in Ref.~\onlinecite{Gottschalk} by introducing a culling hierarchy of orthorhombic boxes with variable axes for the efficient detection of intersections between sets of triangle meshes. While these oriented bounding boxes (OBBs) were found to be particularly well-suited for the case of complex particle models containing large numbers of topological features in close proximity, their efficiency unfortunately rapidly decreases with increasing inter-particle distance, as the relatively high cost of their overlap test progressively outweighs the benefits of their geometric tightness.\cite{Gottschalk}
\par
This limitation is particularly critical in the case of highly anisotropic particles, for which slight angular fluctuations may result in large separation distances between vast sections of the molecules even in the case of small inter-axis separation distances, as illustrated in Fig.~\ref{fig3}a. We therefore propose to combine this hierarchy of OBBs with preliminary overlap queries based on simple bounding sphero-cylinders (SCs) for the efficient determination of neighbour lists in systems of complex particles with high aspect ratios interacting through arbitrary, finite-range atomistic potentials.

\subsection{Construction of the BVH} \label{subsec:Construction of the BVH}
For simplicity, let us consider a generic particle model comprised of $n$ identical atoms, and index their positions by a $3\times n$ coordinate matrix $\mathcal{M} = \{\mathbf{r}_i\}_{i \in \llbracket 1, n \rrbracket}$ expressed in a fixed reference frame. Our algorithm for the recursive construction of the BVH may then be summarised as follows.
\begin{enumerate}[(i), nosep]
\item We determine the molecular long and short axes $\mathbf{u}$ and $\mathbf{v}$ of the particle described by $\mathcal{M}$ through the procedure discussed in the next paragraph, and define its local Cartesian frame as $\mathfrak{R} = (\mathbf{u}, \mathbf{v}, \mathbf{u}\times \mathbf{v})$. 
\item We work out the smallest orthorhombic OBB aligned with $\mathfrak{R}$ that encapsulates all the constituent atoms of $\mathcal{M}$.
\item We then split this OBB along the median plane of its long axis $\mathbf{u}$, and partition the atoms of $\mathcal{M}$ between the two resulting half-boxes; we denote the respective coordinate matrices of their enclosed atoms by $\mathcal{M}_1$ and $\mathcal{M}_2$.
\item We recursively iterate steps (i)--(iii) on the subsystems respectively defined by $\mathcal{M}_1$ and $\mathcal{M}_2$, termed \textit{child nodes} of our bounding hierarchy, until the total number of enclosed atoms at step (i) becomes smaller than an arbitrary number $m$ --- termed the \textit{leaf parameter}.
\item The child nodes at the end of the final recursion step are then referred to as \textit{leaf nodes}, and contain a number $n_l$ of atoms such that $1 \leq n_l \leq m$ and
\begin{equation}
  \sum_{\{l\}} n_l = n,
\end{equation}
where the sum runs over all the leaf nodes $l$ of the hierarchy.
\item We finally append the tightest bounding SC of axis $\mathbf{u}$ enclosing the full particle to the first node of the hierarchy, termed the \textit{root node} of the tree structure, as illustrated in Fig.~\ref{fig3}a.
\end{enumerate}
In order to exploit the pipeline architecture of modern CPUs, atoms contained within the same leaf node are further allocated to contiguous memory locations to maximise the number of cache hits in the tree traversal procedure outlined in Sec.~\ref{subsec:BVHs and binary neighbour search}.\cite{Anderson}

\begin{figure}[htpb]
  \includegraphics[width=\columnwidth]{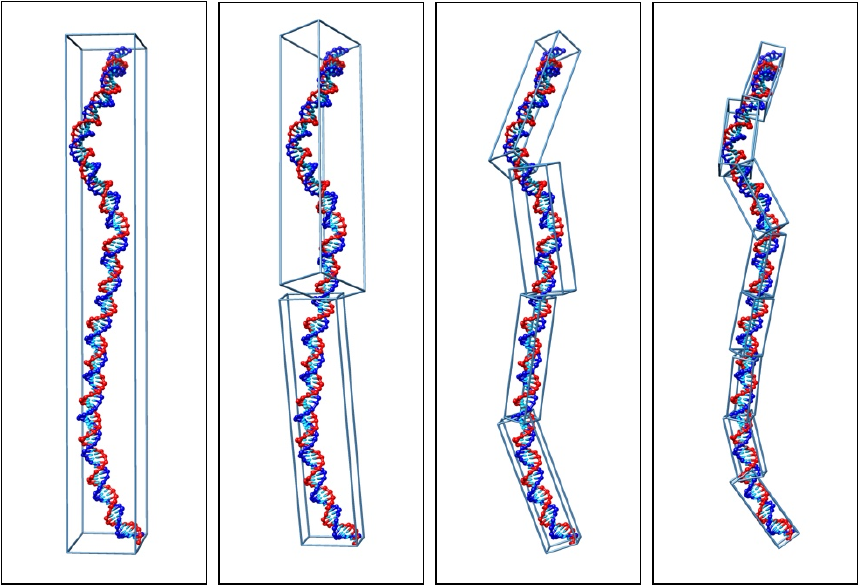}
  \caption{\label{fig1}The first 15 nodes of the culling hierarchy for a 146 base-pair DNA duplex as described by the oxDNA coarse-grained model.\cite{Snodin} The different OBBs are computed using the recursive procedure of Sec.~\ref{subsec:Construction of the BVH}.}
\end{figure}

This geometric construction largely revolves around the determination of the long and short axes $\mathbf{u}$ and $\mathbf{v}$ of the collections of atoms enclosed by the successive nodes, which define the orientations of their corresponding OBBs. A general mathematical method for the efficient calculation of such quantities is known as principal component analysis (PCA), commonly used in data processing applications to determine the directions of extremal statistical dispersion of high-dimensional sets of variables. In the framework of PCA, the long and short axes of a molecule defined by a coordinate matrix $\mathcal{M}$ therefore correspond to the respective eigenvectors associated with the largest and smallest eigenvalues of the covariance matrix of $\mathcal{M}$, assuming the particle center of mass to be set to the origin of the reference frame.
\par
It is easy to prove that this definition for $\mathbf{u}$ is mathematically equivalent to that of Ref.~\onlinecite{Wilson-2} based on the spectral analysis of the molecular inertia tensor, if one assumes all the atoms of the system to bear equal masses. However, a significant advantage of PCA over this previous description lies in its close association with the formalism of singular value decomposition (SVD), as the principal axes of matrix $\mathcal{M}$ may by conveniently worked out as the left-singular vectors of $\mathcal{M}$.\cite{Wall} Therefore, SVD provides a direct numerical route for the evaluation of the local molecular axes as previously defined, and may be performed in $\mathcal{O}(n)$ time for a matrix of size $3\times n$ --- while the computation and spectral decomposition of the inertia tensor would lead to a quadratic overall complexity in the number of enclosed atoms.\cite{Chan} The capabilities of PCA for the efficient fitting of tight bounding volumes have been well-documented in a variety of contexts,\cite{Barequet} and enable our BVH to rapidly converge towards arbitrarily complex molecular shapes, as illustrated in Fig.~\ref{fig1}.
\par
A considerable asset of this object-based acceleration structure is that it only needs to be constructed once per molecule in the case of rigid particle models, whose accessible configuration space may then be fully sampled by uniform rotations and translations of the nodes comprising their initial BVHs. Furthermore, the use of the so-called \textit{median cut algorithm}\cite{Goldsmith} described in step (iii) for the recursive partitioning of the hierarchy leads to the design of a \textit{balanced binary tree} of bounding volumes, in which every non-leaf node is associated with two children nodes enclosing generally comparable numbers of atoms. Such structures have found widespread applications in many areas of computer science owing to their strong synergy with binary search methods, which generally allows for the efficient parsing of complex datasets in logarithmic time.\cite{Bentley} We now proceed to outline the practical implementation of such a scheme for the accelerated computation of inter-molecular interactions.

\subsection{BVHs and binary neighbour search} \label{subsec:BVHs and binary neighbour search}
\begin{figure}[htpb]
  \includegraphics[width=\columnwidth]{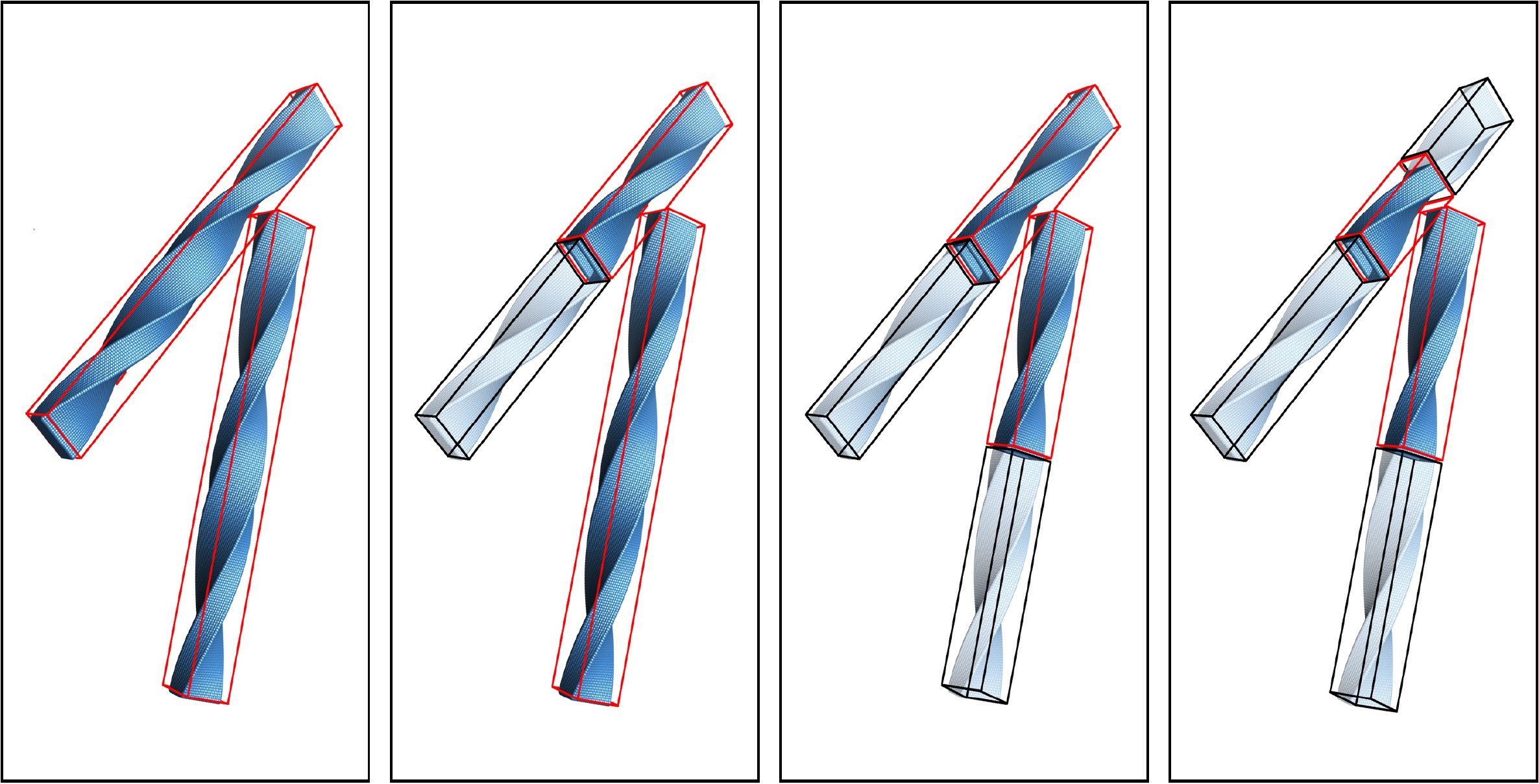}
  \caption{\label{fig2}Schematic representation of the first 4 steps of the BVH-accelerated neighbour search for a pair of twisted cuboidal particles comprised of thousands of fused hard-spheres. Only opaque spheres need to be checked for overlap once the leaf nodes of the hierarchies have been reached.}
\end{figure}

Let us now consider a pair of arbitrary particles defined by coordinate matrices $\mathcal{M}_A$ and $\mathcal{M}_B$, interacting through a pairwise additive interatomic potential $u$ with cutoff range $r_c$. We first construct the respective BVHs of $\mathcal{M}_A$ and $\mathcal{M}_B$ following the procedure of Sec.~\ref{subsec:Construction of the BVH}, and extend all the corresponding bounding volumes by a length $r_c$ along the 3 directions of their local frames. A necessary condition for the existence of non-zero interactions between a given pair of atoms in that case lies in the intersection of all of their respective parent bounding volumes, as illustrated in Fig.~\ref{fig2}.
\par
Based on this simple observation, an efficient neighbour search algorithm may be implemented as follows. 
\begin{enumerate}[(i), nosep]
\item We test the bounding SCs of the root nodes of $\mathcal{M}_A$ and $\mathcal{M}_B$ for mutual overlaps, using the algorithms discussed in the next paragraph. 
\item In the event that they intersect, we proceed to check the corresponding OBBs for overlap. 
\item If these OBBs intersect as well, we recursively iterate steps (ii)--(iii) by descending into the children of the node enclosing the largest number of atoms, until either two leaf nodes are reached or the corresponding OBBs are disjoint.
\item The full interaction energy $U$ of the two molecules is finally given by
\begin{equation}
  U = \sum_{\{l_A, l_B\}} \left (\sum_{i\in l_A} \sum_{j\in l_B} u(\mathbf{r}_i, \mathbf{r}_j) \right ),
\end{equation}
where the outer sum runs over all pairs of overlapping leaf nodes $l_A$ and $l_B$ reached at the end of the traversal process, and the two inner sums over the atoms they respectively enclose.
\end{enumerate}

The overlap tests involving bounding SCs may be efficiently performed using the approach of Vega and Lago,\cite{Vega} which amounts to determining the distance of minimum approach between the axes of their cylindrical backbones, while the intersection checks between OBBs can be expedited by invoking the separating axis theorem as described in Ref.~\onlinecite{Ericson}. Owing to the multiple early exit conditions of the latter test, the relative performance of these two algorithms may vary considerably from case to case; using our implementations, we found the average overlap query times to be roughly 40\% faster for bounding SCs than those of the corresponding OBBs. 
\par
A further advantage of SCs lies in their uniaxial symmetry, which obviates the computation of the full $3\times 3$ orientation matrix $\mathcal{R}$ of the enclosed molecule. In the context of the MC integration of Eqs.~\eqref{eq:elastic_cst} and \eqref{eq:second_virial}, the interaction energy between an arbitrary pair of molecules may thus be efficiently determined by randomly generating their center-of-mass separation vector $\mathbf{r}_{12} \in V$ and unitary main axes $\mathbf{u}_1$, $\mathbf{u}_2$, and restricting the explicit evaluation of $\mathcal{R}_1$ and $\mathcal{R}_2$ to cases in which the corresponding bounding SCs overlap. The performance of the BVH may finally be optimised by fine-tuning the leaf parameter $m$, as illustrated in Fig.~\ref{fig4}; the optimal value for $m$ typically depends on both molecular geometry and the complexity of the underlying inter-atomic potential,\cite{Grudinin-2} and generally needs to be determined empirically for the chosen particle model.

\section{The cholesteric behaviour of hard twisted cuboids} \label{sec:The cholesteric behaviour of hard twisted cuboids}
We now propose to illustrate the application of our method to systems of chiral cuboidal particles obtained by continuously twisting an orthorhombic solid of long axis $\mathbf{u}$ and short axes $\mathbf{v}$, $\mathbf{w}$ through an overall angle $\gamma$ along $\mathbf{u}$, as depicted in Fig.~\ref{fig3}. The biaxiality and chirality of such mesogens may then be independently tuned by respectively varying their transverse cross-section $L_v\times L_w$ and thread angle $\nu$, defined in Fig.~\ref{fig3}b, related to the twist angle $\gamma$ through
\begin{equation}
  \tan \nu = \frac{2L_u}{\gamma \sqrt{L_v^2+L_w^2}},
\end{equation}
with $L_k$ the extent of the cuboids along axis $\mathbf{k}$.
\par
Similarly-shaped particles have been made from DNA using the DNA origami technique,\cite{Rothemund} where low aspect ratio twisted cuboidal origamis have been concatenated to form long particles.\cite{Dietz} A particularly attractive feature of DNA origamis as potential liquid-crystal mesogens lies in the ability to precisely control their molecular shape, twist and curvature,\cite{Dietz} and thus to provide a route towards systematically exploring the link between microscopic features and phase organisation in LCs. Indeed, recently the Dogic group probed the cholesteric behaviour of very high-aspect ratio origamis consisting of 6 parallel double helices in a hexagonal arrangement with varying degrees of twist.\cite{Siavashpouri}
\par
Furthermore, recent experimental evidence suggests that cellulose nano-crystals may naturally adopt configurations with a twisted cuboidal shape,\cite{Cherhal,Usov} which may in turn dictate their chiral nematic arrangement.\cite{Lagerwall-2} Therefore, the theoretical investigation of the cholesteric assembly of hard cuboids as a function of particle twist and biaxiality may provide a valuable starting point for the quantitative understanding of the emergence of chirality from molecular to phase level in such experimental systems.

\subsection{Voxelisation and surface representations} \label{subsec:Voxelisation and surface representations}
\begin{figure}[htpb]
  \includegraphics[width=\columnwidth]{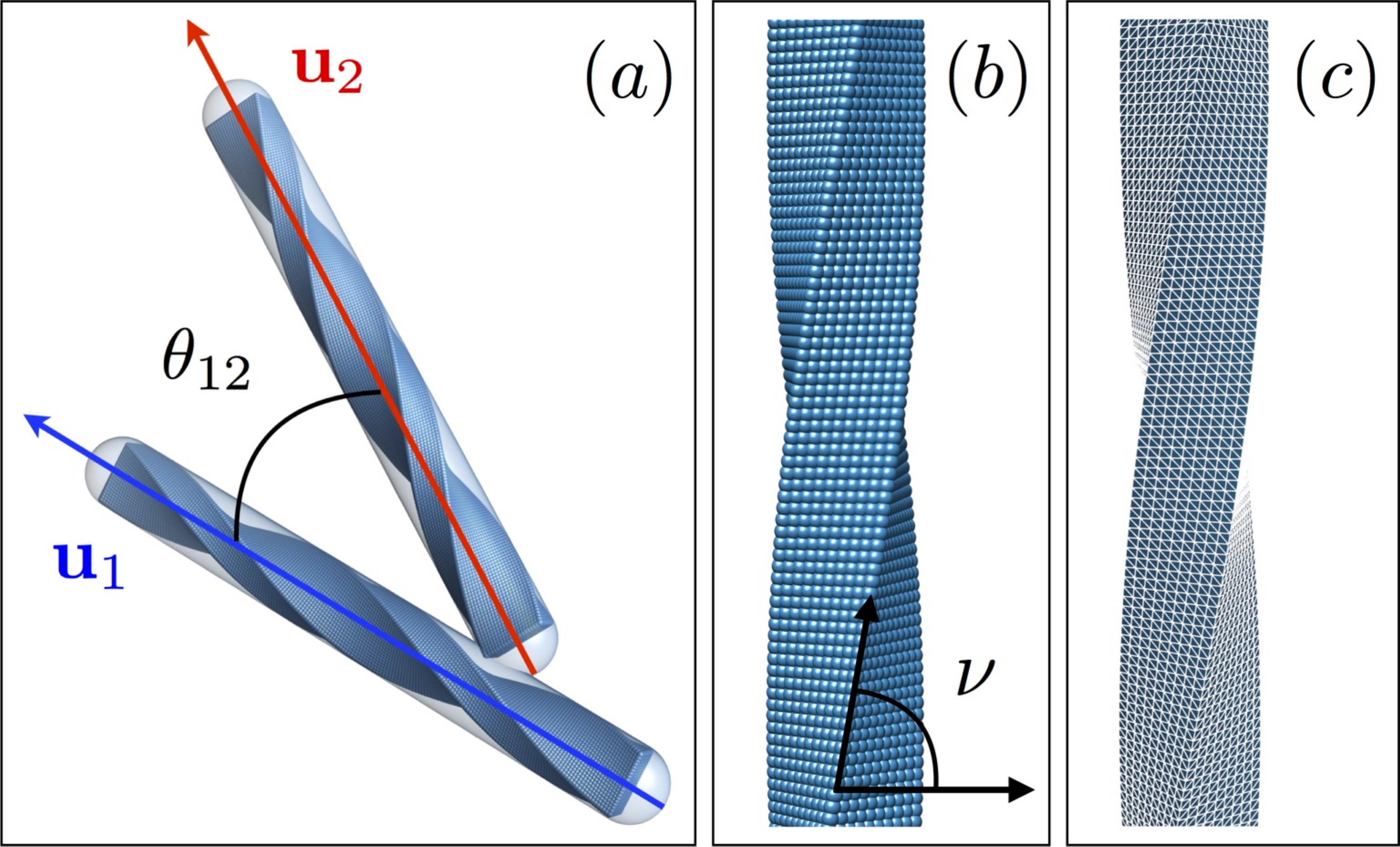}
  \caption{\label{fig3}Long particle axes and local surface representations for hard twisted cuboids. (a) Bounding SCs and inter-axis angle $\theta_{12}$ for a pair of prolate particles. (b) Rasterised model of a hard twisted cuboid; each voxel is represented as the center of a hard sphere of radius $\sigma$. (c) Tesselated mesh obtained by constrained Delaunay triangulation of the former.}
\end{figure}

Despite their conceptual simplicity, the numerical treatment of such particles presents a significant practical challenge owing to the difficulty of accurately describing their detailed structure in terms of simple geometrical primitives. The precise representation of their twist-induced local curvature thus requires the decomposition of their surface into a discrete set of elementary points, termed \textit{voxels}. In this framework, the simplest finite-element representation of a particle may be obtained by associating each voxel with a hard spherical shell of given radius $\sigma$, as illustrated in Fig.~\ref{fig3}b. Denoting by $N_u$, $N_v$ and $N_w$ the respective numbers of voxels uniformly distributed along their 3 orthorhombic axes, the smallest possible shell radius $\sigma_{\rm min}$ is given by
\begin{equation}
  \sigma_{\rm min} = \max_{k\in \{u,v,w\}} \frac{L_k}{N_k-1},
\end{equation}
which ensures that no adjacent spheres are disjoint, and guarantees the integrity of the resulting surfaces.

Overlap queries between such fused hard-sphere models then simply reduce to checking their constituent voxels for intersection, and may thus be efficiently handled by the procedure of Sec.~\ref{sec:Recursive acceleration structures for virial integration} using the simple hard-sphere limit for the interatomic potential $u$,
\begin{align*}
   u(\mathbf{r}_i, \mathbf{r}_j) =
  \begin{dcases}
   +\infty & \text{if } \lVert \mathbf{r}_i - \mathbf{r}_j \rVert \leq \sigma \\
    0  & \text{if } \lVert \mathbf{r}_i - \mathbf{r}_j \rVert > \sigma
  \end{dcases}.
\end{align*}
In the following, we set $\sigma = 1.5\,\sigma_{\rm min}$ and use $N_k = \xi L_k$, with $\xi$ the voxel density per unit length. The total number $n_{\rm vox}$ of voxels comprising a single particle is then
\begin{equation}
  n_{\rm vox} = 2 + \sum_{k \neq l} (N_k-1)(N_l-1),
\end{equation}
where the sum runs over all distinct pairs of axes $k,l\in\{u,v,w\}$.

\par
A limitation of this rasterised representation however lies in the roughness of the resulting surfaces, which entails the use of a high voxel resolution to emulate a smooth particle texture. An alternative geometrical description may be obtained by constrained Delaunay triangulation, which in our case simply amounts to allocating a unique hard triangle to each triplet of neighbouring voxels --- imposing that each voxel is associated to a vertex shared between at most 6 distinct triangles, as depicted in Fig.~\ref{fig3}c. The number $n_s$ of constituent simplices within the resulting mesh is then given by the so-called Euler-Poincar\'e formula with genus 0,\cite{Farkas}
\begin{equation}
  n_s = 2 n_{\rm vox} -4.
\end{equation}
\par
Owing to their ubiquity in computer graphics applications, multiple highly-efficient algorithms have been developed in recent years for the reliable detection of overlaps between such triangle meshes. We here choose to make use of the well-established RAPID open-source library, based on the bounding tree hierarchy introduced in Ref.~\onlinecite{Gottschalk}. This numerical implementation constitutes one of the standard collision detection schemes used in a number of performance-critical applications,\cite{Reggiani} and provides a valuable benchmark for the assessment of the efficiency and accuracy of our approach.
\par
\begin{figure}[htpb]
  \includegraphics[width=\columnwidth]{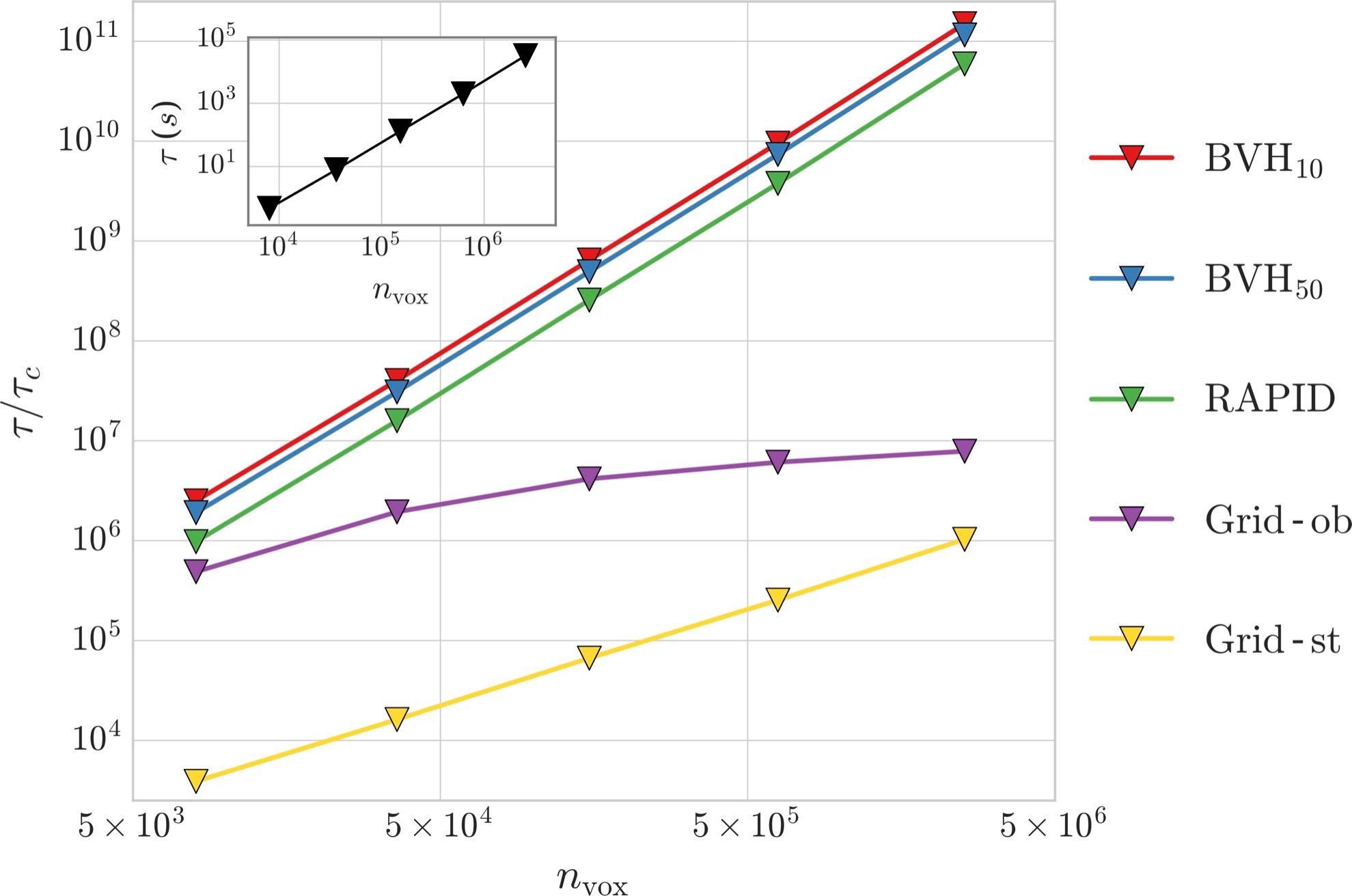}
  \caption{\label{fig4}Relative speedups of the different neighbour search algorithms for hard twisted cuboids with $L_u=100$, $L_v=L_w=1$, $\nu=\SI{80}{\degree}$ and variable numbers of voxels. $\tau$ represents the average time of a naive overlap query between a single pair of particles, plotted in the inset, and $\tau_c$ describes the corresponding numerical costs of the different culling structures described in the text. ${\rm BVH}_m$ denotes the performance of our bounding volume hierarchy using a given value of leaf parameter $m$, c.f.~Sec.~\ref{subsec:BVHs and binary neighbour search}. All measured times were averaged over $10^2$ to $10^5$ randomly generated configurations on an Intel\textsuperscript{\textregistered} Core\texttrademark~i5-4690 CPU.}
\end{figure}

We summarise in Fig.~\ref{fig4} the average processing times of a single overlap test between pairs of twisted cuboids with aspect ratios of 100 and various voxel resolutions using both our approach and the RAPID library, along with standard (st) and object-based (ob) variations of conventional grid acceleration structures. In the latter object-based case, we split the root OBBs of the two particles into two arrays of contiguous bounding boxes aligned with their respective molecular frames, into which constituent voxels may be binned. If the root OBBs intersect, we incrementally traverse the two grids and iterate over the voxels enclosed by each pair of overlapping boxes until a hard sphere collision is detected. We here set the number of boxes allocated along each particle axis $k$ to $n^b_k = N_k/5$, which was found to yield optimal performance for the particles studied, with the corresponding box dimensions $l_k^b$ related to those $l_k$ of the root OBBs through 
\begin{equation*}
  l_k^b = l_k/n^b_k + \sigma \quad \forall k\in\{u,v,w\}. 
\end{equation*}
The standard grid implementation is identical to that described in Ref.~\onlinecite{Grudinin-2} using a cutoff distance of $\sigma$, and is conceptually equivalent to a traditional cell-list approach.

We note that the performance scaling of hierarchy-based methods is found to be sub-logarithmic in the number $n_{\rm vox}$ of elementary voxels per particle, with the average processing time of a pair configuration increasing by roughly 20\% upon raising $n_{\rm vox}$ from 8\,018 to 2\,540\,168, corresponding to a sixteenfold increase of voxel density from $\xi=5$ to $\xi=80$. Accordingly, the associated speedups relative to a naive quadratic implementation --- in which every pair of constituent voxels needs to be checked for overlaps at each step --- range from 6 to 11 orders of magnitude for these systems, with our approach for $m=10$ yielding an increase in performance upwards of 250\% over RAPID on all tests. This difference may be attributed both to the comparatively lower cost of overlap queries between spherical primitives and to our additional use of bounding SCs to efficiently prune non-overlapping configurations, as described in Sec.~\ref{sec:Recursive acceleration structures for virial integration}.
\par
We also remark that the asymptotic scaling of our object-based grid algorithm (Grid-ob in Fig.~\ref{fig4}) is super-linear due to the linear increase of the total number of grid cells with $n_{\rm vox}$, leading to a quadratic worst-case performance in the event that no voxel overlaps are detected after a full grid traversal. Its numerical efficiency however becomes comparable to that of hierarchy-based approaches in the case of highly coarse-grained particles comprised of $n_{\rm vox} \lesssim 500$ voxels, for which computation costs are dominated by voxel overlap queries rather than neighbour search.\cite{Grudinin-2} All culling structures are found to vastly outperform our standard cell-list implementation (Grid-st in Fig.~\ref{fig4}) for the particles studied, despite the linear scaling of the latter.

\subsection{Thread angles and handedness inversion} \label{subsec:Thread angles and handedness inversion}
A further interesting feature of this idealised particle model lies in the continuous threads formed by their uniformly-twisted transverse edges, which constitute a practical realisation of the simplified system of threaded rods first introduced by Straley.\cite{Straley-2} In this context, a twisted cuboid of dimensions $L_u$, $L_v$ and $L_w$ may thus also be considered as a threaded rod of length $L_u$ and groove depth
\begin{equation}
  \label{eq:grooves}
  \Gamma =  \frac{\min\{L_v, L_w\}}{2} \times \left(\sqrt{1 + \chi^2} -1\right),
\end{equation}
where $\chi \equiv \max\{L_v, L_w\} / \min\{L_v, L_w\}$ quantifies the molecular biaxiality of the system.
\par
\begin{figure}[htpb]
  \includegraphics[width=\columnwidth]{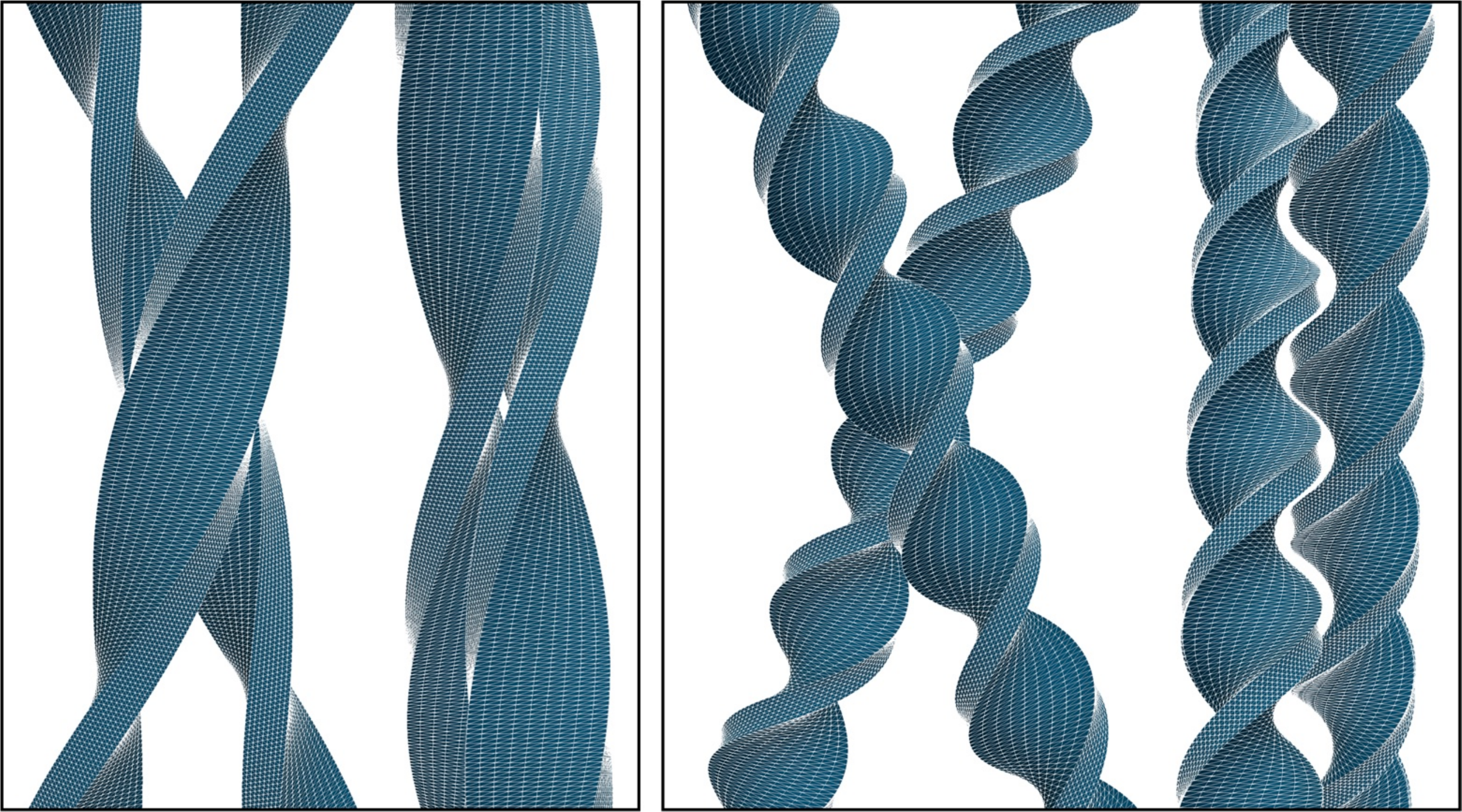}
  \caption{\label{fig5}Front and side views of two-particle configurations at the relative inclination angle $\theta_{12}$ minimising their inter-axis separation distance. The cuboids represented bear a right-handed twist with $L_u=100$, $L_v = 1$, $L_w=3$, $\nu=\SI{70}{\degree}$ (left) and $\nu=\SI{40}{\degree}$ (right). Note the transition of the arrangements from left- to right-handedness with decreasing particle thread angles.} 
\end{figure}

Intriguingly, the handedness of the cholesteric phase formed by such threaded particles has been previously conjectured to depend on the value of their thread angle $\nu$ based on simple geometric arguments.\cite{Straley-2, Lubensky-2, Ferrarini-3} In particular, for $\nu > \SI{45}{\degree}$ the preferred relative twist of two particles at closest contact is of the opposite handedness to the particles, whereas for $\nu < \SI{45}{\degree}$ the twist is of the same handedness, as illustrated in Fig.~\ref{fig5}. At the crossover angle of $\nu = \SI{45}{\degree}$, the particle long axes are expected to lie at a \SI{90}{\degree} angle at their point of closest approach.

However, as is well-known, assuming the particles to adopt such relative inter-axis angles in cholesteric phases leads to the prediction of pitches of the order of just a few particle diameters,\cite{deGennes} suggesting that cholesteric behaviour may not be easily described in terms of single-particle geometric features alone.\cite{Lubensky-1,Lubensky-2} We therefore proceed to investigate the link between cholesteric behaviour and molecular twist to provide a first-hand assessment of the validity of these qualitative assertions. 
\par
\begin{figure}[htpb]
  \includegraphics[width=\columnwidth]{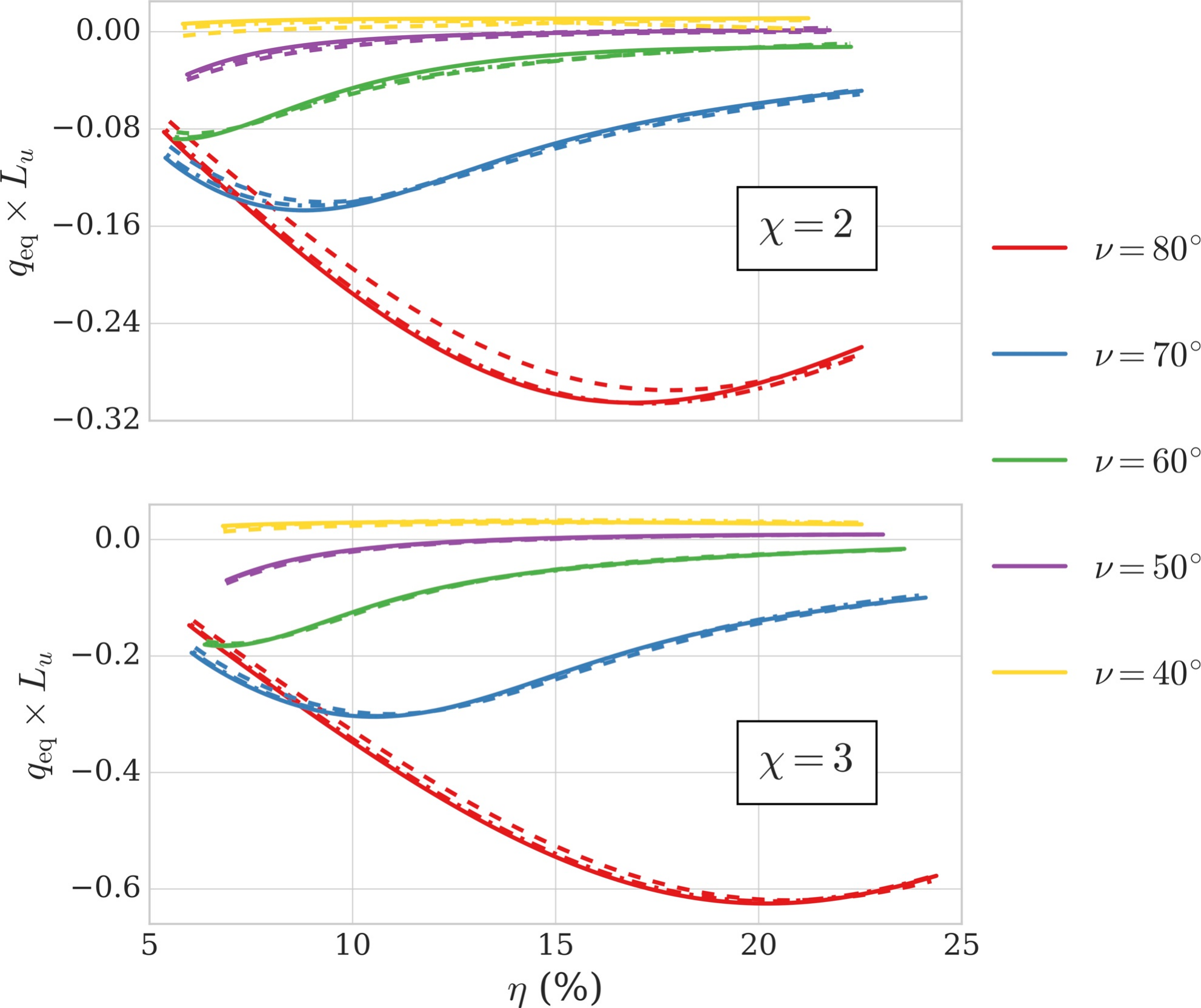}
  \caption{\label{fig6}Cholesteric wavenumber vs.~particle volume fraction for twisted cuboids with $L_u=100$, $L_v = 1$, $L_w = \chi L_v$ and various thread angles $\nu$. Positive values of $q_{\rm eq}$ denote a right-handed cholesteric phase. Dashed and dash-dotted lines respectively denote rasterised particle representations with $\xi=10$ and $\xi=40$ (Fig.~\ref{fig3}b), while solid lines correspond to tesselated surface meshes with $\xi=10$ (Fig.~\ref{fig3}c). All particles possess a right-handed microscopic chirality, and the lowest reported densities correspond to their respective nematic binodals computed following Ref.~\onlinecite{Tortora}.}
\end{figure}

We reproduce in Fig.~\ref{fig6} the concentration dependence of the equilibrium cholesteric wavenumbers $q_{\rm eq}$ of twisted cuboids with various thread angles $\nu$ and rectangular cross sections $\chi$. Firstly, we note that the results are very insensitive to the surface representation of the particles (c.f.~Fig.~\ref{fig3}).  Secondly, consistent with the previous geometric arguments, we find that loosely-threaded particles --- associated with the larger values of $\nu$ --- tend to form cholesteric phases with opposite (left) handedness, while increasing microscopic twist progressively stabilises phases with the same (right) handedness. The crossover region over which this handedness inversion occurs is further found to lie in the range $\nu\in[\SI{40}{\degree}, \SI{50}{\degree}]$, in good apparent agreement with the threshold value $\nu=\SI{45}{\degree}$ inferred from the dense packing of idealised hard threaded rods.\cite{Cherstvy} 

Conversely, we remark that increasing particle twist generally tends to unwind the absolute pitch of the higher-density phases, in stark contrast with intuitive predictions based purely on such geometrical considerations. A simple interpretation of this rather puzzling trend may be obtained by considering the chiral components of the angle-dependent two particle excluded volume,\cite{Wensink-3} formally defined as\cite{Dussi-1, Dussi-2}
\begin{equation}
\begin{split}
  U_{\mathscr{H}}(\theta) = &-\iiint_{\mathscr{H}} d\mathbf{r}_{12} d\mathcal{R}_1 d\mathcal{R}_2 \times f(\mathbf{r}_{12}, \mathcal{R}_1, \mathcal{R}_2) \\ &\times \delta(\cos \theta - \mathbf{u}_1 \cdot \mathbf{u}_2),
\end{split}
\end{equation}
where $\mathscr{H} \in \{\mathscr{L}, \mathscr{R}\}$ denotes the respective contributions of left- and right-handed configurations and the integral runs over the $\mathscr{H}$-handed subset of the two-particle excluded volume manifold. Using this definition, a system of two particles with inter-axis angle $\theta_{12}$ (c.f.~Fig.~\ref{fig3}a) will adopt an entropically-stable right-handed configuration if and only if $U_\mathscr{L}(\theta_{12}) > U_\mathscr{R}(\theta_{12})$. For the sake of normalisation, it is convenient to introduce the dimensionless quantity\cite{Dussi-2}
\begin{equation}
  \Delta_\mathscr{H} U^\ast \equiv \frac{U_\mathscr{L} - U_\mathscr{R}}{U_\mathscr{L} + U_\mathscr{R}} \equiv \frac{\Delta_\mathscr{H} U}{U_\mathscr{L} + U_\mathscr{R}} 
\end{equation}
with $U_\mathscr{L} + U_\mathscr{R}$ the full angle-dependent excluded volume.
\par
\begin{figure}[htpb]
  \includegraphics[width=\columnwidth]{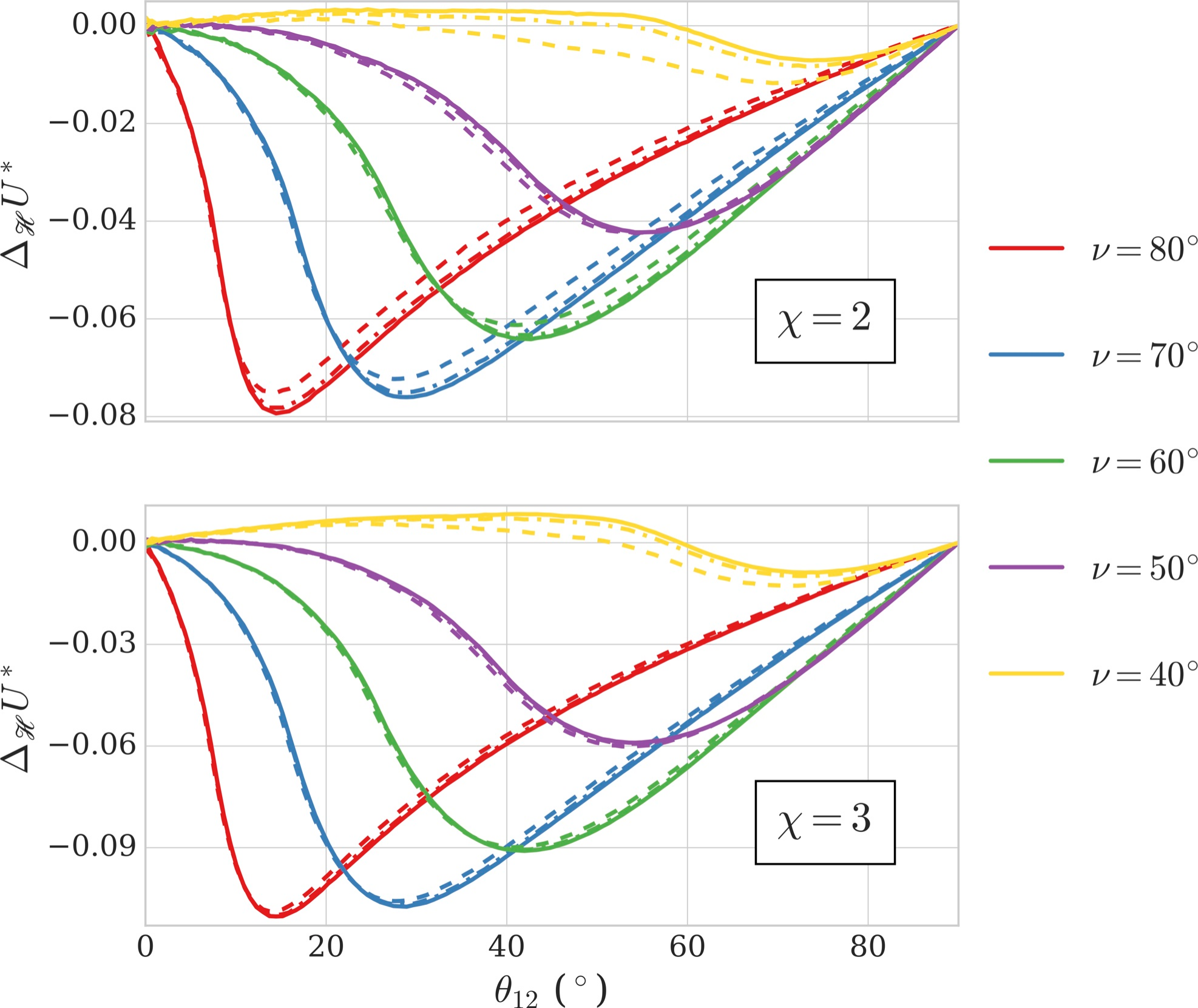}
  \caption{\label{fig7}Normalised two-particle chiral excluded volume as a function of inter-axis angle (c.f.~Fig.~\ref{fig3}a). The $x$-axis is limited to the range $\theta_{12}\in[0,\pi/2]$ owing to the symmetry relation $\Delta_\mathscr{H} U^\ast(\theta_{12}) = -\Delta_\mathscr{H} U^\ast(\pi-\theta_{12})$ imposed by the apolar nature of the particles. Colors and symbols are defined as in Fig.~\ref{fig6}. The global extrema of $\Delta_\mathscr{H} U^\ast$ for tesselated particles with $\chi=3$ and $\nu \in \{\SI{40}{\degree},\SI{70}{\degree}\}$ correspond to the inter-axis angle of their respective configurations of nearest approach as depicted in Fig.~\ref{fig5}.} 
\end{figure}

We reproduce in Fig.~\ref{fig7} the angle dependence of $\Delta_\mathscr{H} U^\ast$ for the systems introduced in Fig.~\ref{fig6}. We note that 
\begin{equation*}
  \Delta_\mathscr{H} U^\ast(\theta_{12}) < 0 \quad \forall \theta_{12}\in[0,\pi/2] 
\end{equation*} 
for all particles with thread angles $\nu \geq \SI{50}{\degree}$, ensuring the formation of a stable left-handed phase. Increasing particle twist progressively moves the location of the extremum of $\Delta_\mathscr{H} U^\ast$ to higher values of $\theta_{12}$, mirroring the larger inter-axis angles required by a pair of such particles to reach their distance of minimal approach. Such configurations are however incompatible with the local orientational alignment inherent to a cholesteric structure, and the particles' sampling of their respective excluded volume manifolds is therefore topologically constrained to the region of small $\abs{\theta_{12}}$, in which $\Delta_\mathscr{H} U^\ast$ progressively vanishes with decreasing thread angles in the range $\nu \in [\SI{50}{\degree}, \SI{80}{\degree}]$. 

This phase-induced restriction of the effective configuration space becomes increasingly stringent at higher particle concentrations and thus prevents the more strongly twisted particles from accommodating their local chirality, thereby accounting for the unwinding of cholesteric pitches observed in this regime. Prior to this, there may also be a regime where phase chirality increases with particle concentration as the effects of short-range chiral interactions become more important. The resulting minimum in the pitch occurs at higher particle concentrations for the less-twisted particles, consistent with the lesser conflict between the preferred twist of the particles and the local nematic ordering. Finally, cholesteric handedness inversions may be attributed to the appearance of positive local extrema of $\Delta_\mathscr{H} U^\ast$ at low values of $\theta_{12}$ for $\nu=\SI{40}{\degree}$, instigating the gradual stabilisation of a right-handed phase.\cite{Wensink-3}
\par
A simple geometric interpretation of these low-angle extrema may be obtained by closer inspection of Fig.~\ref{fig5}. Indeed, it is apparent from the arrangement of twisted cuboids with thread angle $\nu=\SI{70}{\degree}$ that the optimal packing of two such particles may be achieved by locally aligning their respective long faces at their point of close contact. Such configurations may however only be reached in the regime where the width $L_w$ of their long faces is shorter than their microscopic pitch, as increasing the twist of the particles progressively precludes the near proximity of these long faces. In this case, a more entropically-favorable arrangement may rather be obtained by locally aligning the short face of one particle with the long face of the other, as illustrated in the right panel of Fig.~\ref{fig5} for cuboids with $\nu=\SI{40}{\degree}$, leading to comparatively smaller inter-axis angles between adjacent mesogens.
\par
\begin{figure}[htpb]
  \includegraphics[width=\columnwidth]{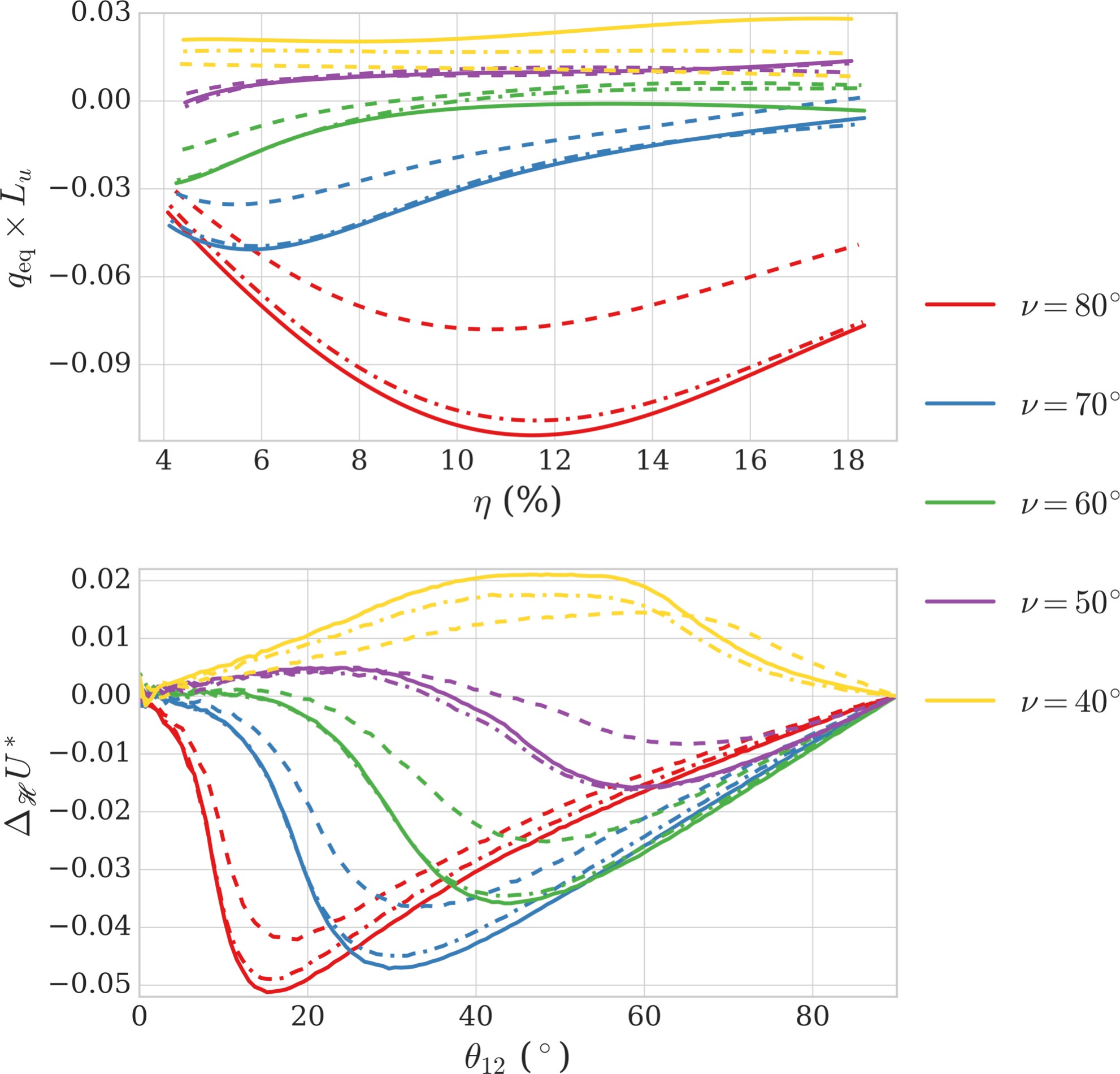}
  \caption{\label{fig8} Top: cholesteric wavenumber as a function of particle volume fraction for twisted cuboids with $L_u=100$, $L_v=L_w=1$ and various thread angles $\nu$. Bottom: normalised two-particle chiral excluded volume as a function of inter-axis angle. Colors and symbols as in Fig.~\ref{fig6}.}
\end{figure}

Lastly, we report in Fig.~\ref{fig8} the density dependence of the equilibrium cholesteric pitch of twisted cuboids with square cross-sections, along with the corresponding angular variations of $\Delta_\mathscr{H} U^\ast$. We note that the shallower grooves of such systems (Eq.~\eqref{eq:grooves}) significantly reduce the chirality of the resulting phases, and increase the sensitivity of their macroscopic pitch to the microscopic details of their surface structure; the results obtained using rasterised particle models nonetheless converge towards those of their smoother tesselated counterparts with increasing $\xi$, as all discretisation schemes lead to equivalent particle descriptions in the limit of high voxel resolutions. Furthermore, the onset of handedness inversions is in this case found to depend very finely on both molecular density and local surface representations, with systems in the range $\SI{50}{\degree} \leq \nu \leq \SI{60}{\degree}$ exhibiting a potential crossover from left- to right-handedness --- associated with a weak small-angle maximum of $\Delta_\mathscr{H} U^\ast$ --- upon slight variations of their concentration and microscopic features.

\begin{figure}[htpb]
  \includegraphics[width=\columnwidth]{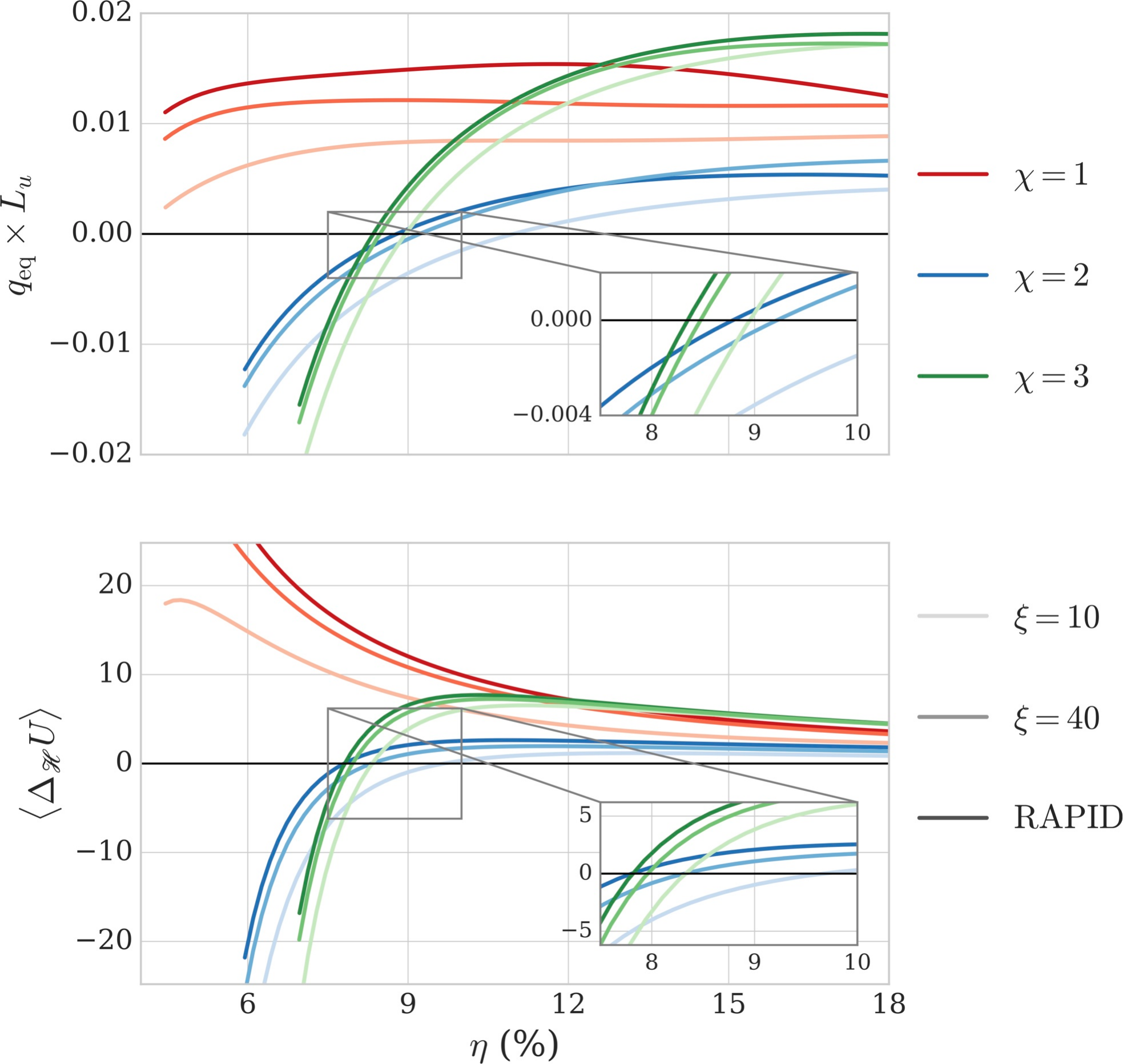}
  \caption{\label{fig9}Cholesteric wavenumber and angle-averaged chiral excluded volume as a function of particle volume fraction for twisted cuboids with $L_u=100$, $L_v = 1$, $L_w = \chi L_v$ and $\nu=\SI{45}{\degree}$. Particle representations are defined as in Fig.~\ref{fig6}, with RAPID denoting tesselated surface meshes with $\xi=10$.}
\end{figure}

Keeping the previous discussion in mind, a more rigorous thermodynamic explanation of these subtle disparities may be achieved by considering the ensemble-averaged chiral excluded volume,\cite{Dussi-1,Dussi-2}
\begin{equation}
  \langle\Delta_\mathscr{H} U\rangle(\rho) = 8\pi^2 \rho \int_0^{\pi/2} d\theta \times \sin \theta \psi_{\rm eq}^{(\rho)}(\theta) \times \Delta_\mathscr{H} U(\theta),
\end{equation}
which combines the purely-geometric quantity $\Delta_\mathscr{H} U$ with the local nematic correlations underlying the uniaxial approximation of our perturbative approach.\cite{Tortora} In this framework, the handedness of cholesteric phases at equilibrium may then be qualitatively captured by the sign of the density-dependent $\langle\Delta_\mathscr{H} U\rangle$, as illustrated in Fig.~\ref{fig9}. The quantitative determination of the magnitude of the corresponding pitches however requires the full evaluation of $\kappa_{11}$ and $\kappa_{01}$, as described in Sec.~\ref{sec:Perturbative density functional theory for cholesteric phases}.

We note that similar handedness inversions were recently reported in cholesteric systems of colloidal hard helices, in which crossovers from left- to right-handedness could be observed upon slight variations of the mesogen shape and concentration.\cite{Ferrarini-4,Dussi-2} A more natural geometric parametrisation for such particles is generally given in terms of their helical pitch $p$ and radius $r$, related to the microscopic thread angle $\nu$ through
\begin{equation}
  \label{eq:helical_thread}
  \nu(r,p) = \arctan \frac{p}{2\pi r}.
\end{equation}
\par
In this case, it was reported in Ref.~\onlinecite{Ferrarini-4} that helices with $\nu \gtrsim \SI{40}{\degree}$ preferably form cholesteric phases of opposite handedness to that of the particles, with lower values of $\nu$ gradually stabilising phases of the same handedness. These trends were independently confirmed by the more extensive numerical analysis of Ref.~\onlinecite{Dussi-2}, which further uncovered a fine dependence of the crossover region on particle aspect ratio and location in the $(r,p)$ parameter plane. It was thus found that helices with large radii $r$ generally undergo a handedness inversion for thread angles $\nu(r,p)\in [\SI{40}{\degree},\SI{60}{\degree}]$, while their shallower-grooved counterparts may exhibit a crossover for values of $\nu$ as large as \SI{75}{\degree} (Eq.~\eqref{eq:helical_thread}).\cite{Dussi-2}
\par
This qualitative trend seemingly mirrors the shift of the handedness inversion range to higher values of $\nu$ observed in Fig.~\ref{fig8} for twisted cuboids with $\chi=1$, associated with smaller groove depths. However, the focus of Refs.~\onlinecite{Ferrarini-4} and~\onlinecite{Dussi-2} on particles with much shorter aspect ratios than considered here precludes the further quantitative comparison of our results with the cholesteric behavior of hard helices. It should nonetheless be noted that the shortest cholesteric pitches observed here, amounting to roughly 1000 particle diameters, are about one order of magnitude larger than those reported in these previous studies, and are comparatively closer to the typical values observed in experimental cholesteric systems.\cite{Ferrarini-3} These tighter pitches may be typically obtained for $\nu \sim \SI{65}{\degree}$ in the case of hard helices and $\nu \in [\SI{70}{\degree},\SI{80}{\degree}]$ for twisted cuboids, highlighting the complex relationship between particle twist and phase chirality in both systems.
\par
Finally, it is worth remarking that although the qualitative shape of the different curves at given angle $\nu$ does not differ much with varying $\chi$, the pitch of the corresponding phases tends to significantly tighten with increasing cross-section anisotropy. While this effect seemingly concurs with the numerical results of Ref.~\onlinecite{Dussi-3} on simple model systems of twisted triangular prisms, we recall that the assumption of local phase uniaxiality underlying Eq.~\eqref{eq:density_onsager} precludes the treatment of long-range biaxial correlations in our current density-functional framework, and may thus become increasingly inadequate for higher values of $\chi$.\cite{Cuestos} This basic symmetry constraint may therefore lead to growing inaccuracies in the description of the corresponding equilibrium cholesteric properties, and could account for some of the discrepancies reported in Ref.~\onlinecite{Dussi-3} for the theoretical predictions of pitch dependence on particle biaxiality. The relaxation of local cylindrical invariance would allow for a direct investigation of these considerations, and may lead to a more realistic representation of phase structure in these cases.\cite{Tortora}

\section{Conclusion and outlook} \label{sec:Conclusion and outlook}
We have introduced in detail a novel numerical virial integration scheme for highly anisotropic particles with arbitrarily complex molecular structures. Its implementation within an efficient perturbative density-functional framework enables us to probe the cholesteric behaviour of twisted cuboidal mesogens with high surface resolutions, and to provide a quantitative assessment of the intricate link between microscopic twist and cholesteric assembly in the case of hard particles with continuous threads. We thus report that deep-grooved systems undergo a handedness inversion for thread angles in the vicinity of the previously-proposed phenomenological value $\nu=\SI{45}{\degree}$, while the behaviour of their shallower counterparts depends on the fine details of their surface representation. The relationship between helical twisting power and particle structure is found to be non-trivial in both cases, and may be qualitatively explained in terms of subtle angular variations of the chiral component of their mutual excluded volumes. 
\par
The performance of our hierarchy-based approach for the determination of their hard overlaps is shown to be several orders of magnitude faster than traditional cell lists, and nearly three times as fast as dedicated collision detection libraries in the case of twisted cuboids with high aspect ratios. Even though the polygon mesh support of the latter admittedly provides for a more natural description of such hard faceted particles, the strength of our method lies in its applicability to a much wider variety of finite-range force fields and its simple generalizability to polymolecular systems and multiple interaction potentials --- e.g.~by constructing one BVH per particle and interaction type.\cite{Glotzer-1}
\par
While we have here largely focused on the applications of BVHs to the direct determination of non-bonded energy terms involving pairs of rigid mesogens, we note that the procedure described in Sec.~\ref{sec:Recursive acceleration structures for virial integration} may also be more broadly applied to the computation of Verlet lists by introducing a finite Verlet skin as an additional BVH padding parameter.\cite{Allen} The superior geometric tightness of such object decomposition methods greatly reduces the number of spurious distance calculations ensuing from the use of standard cell lists, which constitute one of the recurrent computational bottlenecks of molecular simulations, and could thus provide sizeable performance gains for the determination of neighbour lists in a number of contexts.\cite{Grudinin-1,Grudinin-2,Glotzer-2}
\par
However, the top-down construction scheme adopted in Sec.~\ref{subsec:Construction of the BVH} may in that case not be ideally suited to modern computing architectures, owing to the limited parallelism of its root-based tree initialization process. These considerations are largely irrelevant for the virial-type calculations considered here, which are rather parallelised at the level of the Monte-Carlo integrator to process multiple configurations concurrently, but may entail the use of more involved bottom-up algorithms to efficiently build BVHs in the framework of general-purpose molecular simulations.\cite{Glotzer-1,Glotzer-2}
\par
An interesting further application of BVHs in our case stems from the seminal work of Fynewever and Yethiraj,\cite{Fynewever} who proposed to account for intra-molecular mechanics in the formalism of DFT by introducing the ensemble average of the two-particle excluded volume Eq.~\eqref{eq:second_virial}. The underlying conformational integrals may here be conveniently performed within the numerical procedure of Sec.~\ref{sec:Perturbative density functional theory for cholesteric phases} by randomly parsing trajectories of representative single-particle configurations --- as generated by a relevant molecular model --- in the MC computation of virial integral Eqs.~\eqref{eq:elastic_cst} and \eqref{eq:second_virial}.\cite{Tortora} In this framework, energy calculations may be efficiently handled by building a forest of bounding hierarchies, in which a single tree of bounding structures is constructed for every configuration of the trajectory file, and by applying the traversal routines of Sec.~\ref{subsec:BVHs and binary neighbour search} to a random pair of trees at every MC integration step.
\par
The combined use of BVHs and DFT thus opens up a performant route to investigate the cholesteric and nematic behaviour of vast range of particle models, taking into account the effects of both inter- and intra-molecular mechanics. The implementation of optimised methods for the general evaluation of virial integrals and Frank elastic constants allows us to tackle molecular descriptions with a considerably higher degree of complexity than those considered in previous numerical and analytical efforts.\cite{Ferrarini-1, Ferrarini-2, Ferrarini-4, Dussi-1,Dussi-2} The natural suitability of our theoretical description to the regime of high mesogen anisotropy and weak phase chirality further enables us to tackle a wide variety of experimentally-relevant systems which may not be easily probed using traditional molecular simulations. For instance, tests have shown that we are able to apply the current approach to address the cholesteric assembly of the DNA origamis studied in Ref.~\onlinecite{Siavashpouri} using a description that explicitly represents each of the ${\sim}15\,000$ nucleotides in the origami. In this density-functional framework, the microscopic investigation of cholesteric behaviour is therefore no longer limited by analytical constraints or computational capabilities, but rather by the quality of available particle models and the accuracy of the fundamental DFT assumptions.\cite{Tortora}
\par
A potential shortcoming of the latter lies in the inability of DFT to reproduce the scaling behaviour of highly flexible polymers,\cite{Binder-1} which may restrict its applicability to systems characterised by large persistence lengths. In this context, the thorough assessment of the validity of theoretical predictions requires their extensive comparison with the results of molecular simulations obtained using identical microscopic Hamiltonian descriptions.\cite{Binder-2} The application of our method to nematic phases of semi-flexible chains with various bending rigidities would obviate the numerical approximations introduced by previous studies for the derivation of the corresponding free energies,\cite{Khokhlov, Odijk-1, Chen} and allow for the thorough appraisal of the Onsager DFT in these cases.

\begin{acknowledgments}
This project has received funding from the European Union's Horizon 2020 research and innovation programme under the Marie Sk\l{}odowska-Curie Grant Agreement No.~641839. The authors would like to acknowledge the use of the University of Oxford Advanced Research Computing (ARC) facility in carrying out this work. http://dx.doi.org/10.5281/zenodo.22558. We are grateful to the UK Materials and Molecular Modelling Hub for computational resources, which is partially funded by EPSRC (EP/P020194/1).
\end{acknowledgments}


\bibliography{jcp}


\end{document}